\documentclass[%
 reprint,
superscriptaddress,
showpacs,
 amsmath,amssymb,
 aps,
prl,
floatfix,
]{revtex4-1}

\usepackage{dcolumn}% Align table columns on decimal point
\usepackage{bm}% bold math

\usepackage{graphicx} % Include figure files
\usepackage{color}

\usepackage{ulem} % sout

\usepackage[dvipdfmx,
colorlinks=true, allcolors=blue,
setpagesize=false, bookmarks=false
]{hyperref}

\begin{document}
%\title{Pump-probe spectroscopy of the one-dimensional Mott insulator: excitations via photon absorption and quantum tunneling}
\title{Glassy dynamics of the one-dimensional Mott insulator excited by a strong terahertz pulse}
\author{Kazuya Shinjo}
%\email{shinjo@rs.tus.ac.jp}
\affiliation{Department of Applied Physics, Tokyo University of Science, Tokyo 125-8585, Japan}
\affiliation{Computational Quantum Matter Research Team, RIKEN Center for Emergent Matter Science (CEMS), Wako, Saitama 351-0198, Japan}
\author{Shigetoshi Sota}
\affiliation{Computational Materials Science Research Team,
RIKEN Center for Computational Science (R-CCS), Kobe, Hyogo 650-0047, Japan}
%\email{shinjo@rs.tus.ac.jp}
\author{Takami Tohyama}
%\email{tohyama@rs.tus.ac.jp}
\affiliation{Department of Applied Physics, Tokyo University of Science, Tokyo 125-8585, Japan}

\date{\today}% It is always \today, today,
             
%\pacs{****}

\begin{abstract}
The elucidation of nonequilibrium states in strongly correlated systems holds the key to emergence of novel quantum phases. The nonequilibrium-induced insulator-to-metal transition is particularly interesting since it reflects the fundamental nature of competition between itinerancy and localization of the charge degrees of freedom. We investigate pulse-excited insulator-to-metal transition of the half-filled one-dimensional extended Hubbard model. Calculating the time-dependent optical conductivity with the time-dependent density-matrix renormalization group, we find that strong mono- and half-cycle pulses inducing quantum tunneling strongly suppress spectral weights contributing to the Drude weight $\sigma_\text{D}$, even if we introduce a large number of carriers $\Delta n_\text{d}$. This is in contrast to a metallic behavior of $\sigma_\text{D}\propto \Delta n_\text{d}$ induced by photon absorption and chemical doping. The strong suppression of $\sigma_\text{D}$ in quantum tunneling is a result of the emergence of the Hilbert-space fragmentation, which makes pulse-excited states glassy.
\end{abstract}
\maketitle

%%%%%%%%%%%%%%%%%
%
%%%%%%%%%%%%%%%%%
% ki
\textit{Introduction.}~The elucidation of nonequilibrium states in strongly correlated systems is of great interest since it promises to open a door to the emergence of novel quantum phases.
Nonequilibrium quantum many-body states have recently been investigated not only in solids with light and electric fields~\cite{Yu1991, Taguchi2000, Iwai2003, Cavalleri2004, Okamoto2007, Takahashi2008, Al-Hassanieh2008, Wall2011, Okamoto2011, Liu2012, Yamakawa2017, Ishihara2019} but also in trapped ions~\cite{Blatt2012, Monroe2021}, cold atoms~\cite{Eisert2015, Bernien2017, Senaratne2018}, and quantum circuits~\cite{Martinez2016, Lamm2018, Smith2019, Lin2021, Benedetti2021, Mi2022}.
One of the most significant challenges in this field is how to preserve nonequilibrium states, such as the Floquet states~\cite{Bukov2015, Eckardt2015, Oka2019}, from thermalization~\cite{Dalessio2016, Deutsch1991, Srednicki1994, Rigol2008}, for which the realization of many-body localization (MBL)~\cite{Nandkishore2015, Altman2015, Abanin2019} may hold the key.
% sho
Also, the nonequilibrium-induced insulator-to-metal transition is a fundamental issue associated with competition between itinerancy and localization of charge degrees of freedom.
The photoinduced insulator-to-metal transitions~\cite{Taguchi2000, Iwai2003, Okamoto2007, Takahashi2008, Al-Hassanieh2008, Wall2011} due to photon absorption have been suggested in the one-dimensional (1D) Mott insulator.
Similarly, non-absorbable terahertz photons with strong intensity have been suggested to induce a metallic state~\cite{Liu2012, Yamakawa2017} via quantum tunneling~\cite{Oka2003, Oka2005, Oka2008, Oka2010, Eckstein2010, Oka2012}.

% ten
Until now it has been commonly accepted that the breakdown of the Mott insulators via electric pulses leads to metallic states.
However, we raise question about the validity of this understanding.
To answer this question, we examine the possibility of the emergence of novel quantum phases such as glass phases with intermediate properties between itinerancy and MBL.

% ketsu
In this Letter, we investigate pulse-excited states of the half-filled 1D extended Hubbard model (1DEHM) using the time-dependent density-matrix renormalization group (tDMRG)~\cite{White1992, White2004, Daley2004}.
We propose a Mott transition to glassy states induced by mono- and half-cycle terahertz pulses.
If we excite the Mott insulating state via photon absorption, we obtain metallic states with large spectral weights contributing to the Drude component $\sigma_\text{D}$.
In contrast, we find that strong electric fields inducing the Zener breakdown~\cite{Zener1934} strongly suppress $\sigma_\text{D}$, even if we introduce a large number of carriers.
We consider that the emergence of the Hilbert-space fragmentation~\cite{Moudgalya2019, Rakovszky2020, Khemani2020, Sala2020, Scherg2021, Desaules2021, Herviou2021, Kohlert2021, Papic2021, Moudgalya2021} due to high fields leads to glassy dynamics~\cite{Amir2011, Gopalakrishnan2014, vanHorssen2015, Prem2017, Pretko2017c, Lan2018} as seen in fracton systems~\cite{Pretko2017a, Pretko2017b, Pretko2018, Williamson2019, Sous2020a, Sous2020b, Pretko2020, Nandkishore2019}.

\textit{Model and method.}~To investigate nonequilibrium properties of the 1D Mott insulator, we use 1DEHM with a vector potential $A(t)$ defined as 
\begin{align}\label{Eq-Hamiltonian}
\mathcal{H}=&-t_\mathrm{h}\sum_{i,\sigma} B_{i,\sigma}
+ U\sum_i n_{i,\uparrow}n_{i,\downarrow} + V\sum_i n_{i}n_{i+1},
\end{align}
where $B_{i,\sigma}=e^{iA(t)} c_{i,\sigma}^\dag c_{i+1,\sigma}+\text{H.c.}$, $c_{i,\sigma}^{\dag}$ is the creation operator of an electron with spin $\sigma (= \uparrow, \downarrow)$ at site $i$, and $n_i=\sum_\sigma n_{i,\sigma}$ with $n_{i,\sigma}=c^\dagger_{i,\sigma}c_{i,\sigma}$.
We consider $(U,V)=(10,3)$ taking the nearest-neighbor hopping $t_\mathrm{h}$ to be the unit of energy ($t_\mathrm{h}=1$), which describes the optical conductivity in a 1D Mott insulator ET-F$_{2}$TCNQ~\cite{Yamaguchi2021}.
Spatially homogeneous electric field $E(t)=-\partial_{t}A(t)$ applied along the chain is incorporated via the Peierls substitution in the hopping terms~\cite{Peierls1933}.
Unless otherwise noted, we consider the half-filled 1DEHM with $L=32$ sites.
Note that we set the light velocity $c$, the elementary charge $e$, the Dirac constant $\hbar$, and the lattice constant to 1.

\begin{figure*}[t]
  \centering
    \includegraphics[clip, width=40pc]{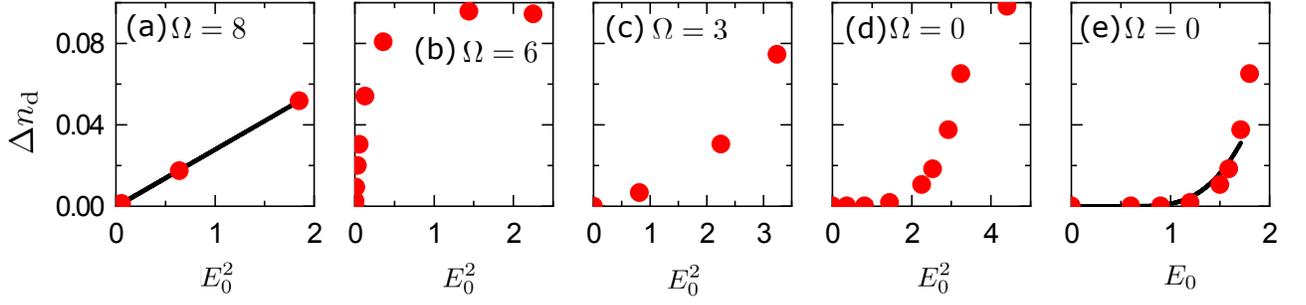}
    \caption{$\Delta n_\text{d}$ of the $L=32$ half-filled 1DEHM for $(U,V)=(10,3)$ excited by electric pulses. Red points are $\Delta n_\text{d}$ as a function of $E_{0}^{2}$ for (a) $\Omega=8$ with a black line for eye guide, (b) $\Omega=6$, (c) $\Omega=3$, and (d) $\Omega=0$. (e) Red points are $\Delta n_\text{d}$ as a function of $E_{0}$ for $\Omega=0$. The black line shows a fitted curve proportional to $E_{0}\exp \left(-\pi E_\text{th}/E_{0} \right)$.}
    \label{Fig1}
\end{figure*}

We assume that pulses have time dependence determined by $A(t)=A_\text{pump}(t)+A_\text{probe}(t)$ with $A_\text{probe}(t)=A_0^\text{pr} e^{-\left(t-t_0^\text{pr}\right)^2/\left[2(t_\mathrm{d}^\text{pr})^2\right]} \cos \left[\Omega^\text{pr}(t-t_0^\text{pr})\right]$ for probe pulses.
Unless otherwise noted, we use $A_\text{pump}(t)=A_0 e^{-(t-t_0)^2/(2t_\mathrm{d}^2)} \cos\left[\Omega(t-t_0)\right]$ for pump pulses.
We set $A_0^\text{pr}=0.001$, $\Omega^\text{pr}=10$, $t_\text{d}^\text{pr}=0.02$, and $t_{0}^\text{pr}=t_{0}+\tau$, where $\tau$ indicates the delay time between pump and probe pulses.
We obtain time-dependent wave functions by the tDMRG implemented by the Legendre polynomical~\cite{Shinjo2021, Shinjo2021b} employing open boundary conditions and keep $\chi=3000$ density-matrix eigenstates.
We obtain both singular and regular parts of the optical conductivity in nonequilibrium $\sigma(\omega,\tau) = \frac{j_\text{probe}(\omega,\tau) }{i(\omega +i\gamma)LA_\text{probe}(\omega)}$~\cite{Shao2016, Shinjo2018, Rincon2021}, where $A_{\text{probe}}(\omega)$ and $j_\text{probe}(\omega,\tau)$ are the Fourier transform of $A_\text{probe}(t)$ and current induced by a probe pulse, respectively (see Sec.~S1 in the Supplemental Material~\cite{Supplement}).
$\gamma$ indicates a broadening factor.

%%%%%%%%%%%%%%
% Fig1
%%%%%%%%%%%%%%

\textit{Doublon density.}~First of all, we demonstrate how pumping energy makes a difference in carrier production. 
Figure~\ref{Fig1} shows how much electric pulses with $(t_\text{d},t_{0})=(2,10)$ change doublon density $\Delta n_\text{d}=\frac{1}{L}\bigl[ \overline{\langle I \rangle_{t}} - \langle I\rangle_0 \bigr]$ in 1DEHM, where $I=\sum_{j}n_{j,\uparrow}n_{j,\downarrow}$, $ \overline{\langle \mathcal{O} \rangle_{t}}$ is the average of an expectation value of an operator $\mathcal{O}$ from $t=21$ to $22$ just before a probe pulse is applied, and $\langle \mathcal{O} \rangle_0$ is an expectation value of $\mathcal{O}$ for a ground state.
We focus on $\Delta n_\text{d}<0.1$, which can be achieved with experiments.
$\Delta n_\text{d}$ oscillates even after pulse decay, but their amplitudes are smaller than the radius of red points in Fig.~\ref{Fig1}.
Since Re$[\sigma(\omega,\tau<0)]$ has an excitonic level at $\omega=\omega_1$ and a continuum begins at $\omega=\omega_\text{c}$~\cite{Stephan1996, Gebhard1997, Essler2001, Jeckelmann2003, Yamaguchi2021}, where $(\omega_{1},\omega_\text{c})=(6,6.5)$ for $(U,V)=(10,3)$, a pump pulse with $\Omega=8$ excite electrons in a continuum leading to $\Delta n_\text{d}\propto E_{0}^{2}$ [see Fig.\ref{Fig1}(a)] as discussed in Ref.~\cite{Oka2012} with the amplitude of electric fields $E_{0}$.
Taking $\Omega=\omega_{1}$, we can efficiently excite doublons and holons even for small $E_{0}$ [see Fig.~\ref{Fig1}(b)].
For subgap excitations, i.e., $\Omega<\omega_{1}$, electrons are excited by a nonlinear process, which is classified into multiphoton absorption and  quantum tunneling.
The crossover between them is called the Keldysh crossover~\cite{Keldysh1965}.
Figures~\ref{Fig1}(c) and \ref{Fig1}(d) show $\Delta n_\text{d}$ generated by two-photon absorption and quantum tunneling, respectively.
For $\Omega=0$ mono-cycle pulses, we find that $\Delta n_\text{d}$ follows a threshold behavior $\Delta n_\text{d} \propto E_{0}\exp \left(-\pi E_\text{th}/E_{0} \right)$~\cite{Oka2012} as indicated by the black line in Fig.~\ref{Fig1}(e).
Using this relation, we can estimate the doublon-holon correlation length $\xi \simeq \omega_{1}/(2E_\text{th}) \sim 1.5$.

%%%%%%%%%%%%%%
% Fig2
%%%%%%%%%%%%%%
\begin{figure}[t]
  \centering
    \includegraphics[clip, width=20pc]{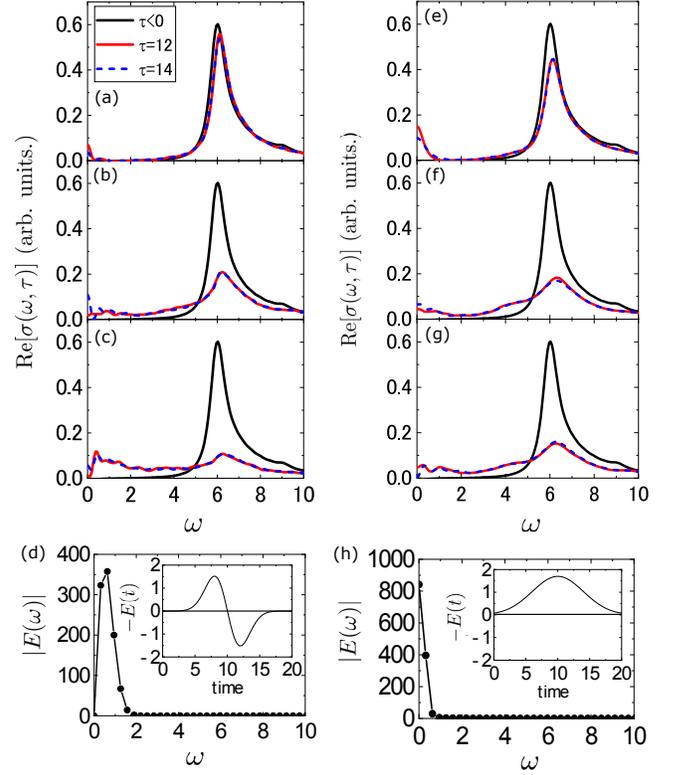}
    \caption{Re$[\sigma(\omega,\tau)]$ excited by $\Omega=0$ mono- [half-]cycle pulses for $t_\text{d}=2$ [$t_\text{d}=4$] with (a) $E_{0}=1.5$, (b) $E_{0}=1.8$, and (c) $E_{0}=2.1$ [(e) $E_{0}=1.7$, (f) $E_{0}=1.9$, and (g) $E_{0}=2.0$]. Black, red, and blue-dashed lines are for $\tau<0$, $\tau=12$, and $14$, respectively. (d) [(h)] $|E(\omega)|$ of a mono- [half-]cycle pulse with $E_{0}=1.5$ [$E_{0}=1.7$]. The inset indicates $-E(t)$. (a)--(c) and (e)--(g) are obtained with the half-filled $L=32$ 1DEHM for $(U,V)=(10,3)$ taking $\gamma=0.4$.}
    \label{Fig2}
\end{figure}
\textit{Glassy dynamics.}~We show in Fig.~\ref{Fig2} the results of 1DEHM excited by a quantum tunneling with strong $\Omega=0$ pulses whose energy is in terahertz band.
We show Re$[\sigma(\omega,\tau)]$ excited by mono-cycle pulses with $(\Omega,t_\text{d})=(0,2)$ for various $E_{0}$ in Figs.~\ref{Fig2}(a)-\ref{Fig2}(c).
$|E(\omega)|=\left|\int dt e^{i\omega t}E(t)\right|$ with $E_{0}=1.5$ shown in Fig.~\ref{Fig2}(d) indicates that the photon energy is too small to excite the Mott gap.
We obtain $\Delta n_\text{d}=0.01$, 0.07, and 0.1 for Figs.~\ref{Fig2}(a), \ref{Fig2}(b), and \ref{Fig2}(c), respectively.
The spectral weights above the Mott gap transfer to lower energies, but we find that the Drude weight $\sigma_\text{D}$, which we define as spectral weight below $\omega=0.15$ (see Secs.~S2 and S3 in the Supplemental Material~\cite{Supplement}), is not proportional to $\Delta n_\text{d}$ but is strongly suppressed even if we take large $\Delta n_\text{d}$ as shown in Figs.~\ref{Fig2}(b) and \ref{Fig2}(c).
Note that the Drude weight appears at $\omega \neq 0$ due to a finite-size effect and its peak approaches $\omega=0$ as $L$ increases~\cite{Hashimoto2016, Shao2019, Shinjo2021}. 
For $L=32$, we can mask this finite-size effect by taking $\gamma = 0.4$.
For half-cycle pulses, we obtain Re$[\sigma(\omega,\tau)]$ as shown in Figs.~\ref{Fig2}(e)-\ref{Fig2}(g).
$|E(\omega)|$ given by $E(t)=E_0 e^{-(t-t_0)^2/(2t_\mathrm{d}^2)} \cos[\Omega(t-t_0)]$ for $(\Omega,t_\text{d})=(0,4)$ is shown in Fig.~\ref{Fig2}(h).
We obtain $\Delta n_\text{d}=0.02$, 0.07, and 0.08 for Figs.~\ref{Fig2}(e), \ref{Fig2}(f), and \ref{Fig2}(g), respectively.
Even if we find finite $\sigma_\text{D}$ as shown in Fig.~\ref{Fig2}(e) with small $\Delta n_\text{d}$, further increase in $\Delta n_\text{d}$ does not enhance $\sigma_\text{D}$ as shown in Figs.~\ref{Fig2}(f) and \ref{Fig2}(g), but rather suppresses it.

The strong suppression of $\sigma_\text{D}$ suggests that strong fields localize nonequilibrium states.
When a thermal state with $\sigma_\text{D}\neq0$ approaches an MBL state with $\sigma_\text{D}=0$, $\sigma_\text{D}$ is suppressed and the center of gravity of low-energy spectral weights shifts to higher energy~\cite{Barisic2010, Gopalakrishnan2015, Steinigeweg2016}, which is similar to the structure seen in Figs.~\ref{Fig2}(b), \ref{Fig2}(c), \ref{Fig2}(f), and \ref{Fig2}(g) when $E_{0}$ is large.

%%%%%%%%%%%%%%
% Fig3
%%%%%%%%%%%%%%
\begin{figure}[t]
  \centering
    \includegraphics[clip, width=20pc]{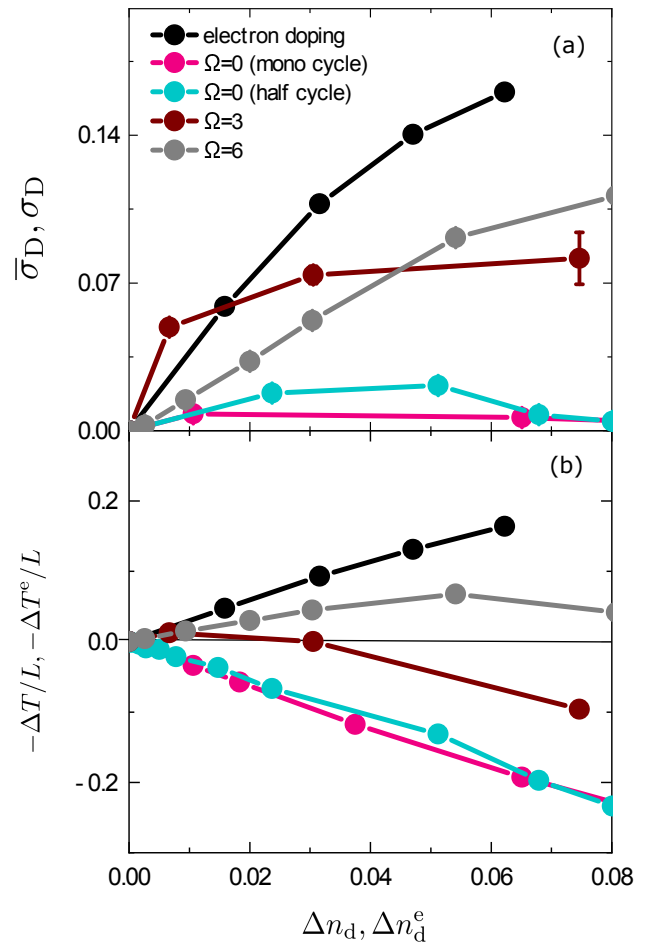}
    \caption{(a) $\overline{\sigma}_\text{D}$ as the function of $\Delta n_\text{d}$ and $\Delta n_\text{d}^\text{e}$. $\gamma=0.4$ is taken. (b) $-\Delta T/L$ as the function of $\Delta n_\text{d}$ and $-\Delta T^\text{e}/L$ as the function of $\Delta n_\text{d}^\text{e}$. All plots are obtained for the $L=32$ 1DEHM.}
    \label{Fig3}
\end{figure}
The suppression of the Drude weight is clearly shown in Fig.~\ref{Fig3}(a) if we compare $\overline{\sigma}_\text{D}$ (see below) induced by $\Omega=0$ pulses (see magenta and light blue points) with those by photon absorption with $\Omega=3$ (see brown points) and $\Omega=6$ (see gray points) pulses as well as electron doping (see black points).
Here, we introduce an time-averaged Drude weight $\overline{\sigma}_\text{D}=\frac{1}{2}\sum_{\tau=12,14}\int_{\omega=0}^{2\eta}d\omega \text{Re}\sigma(\omega,\tau)$ in Fig.~\ref{Fig3}(a) with $2\eta=0.15$. 
Note that carrier density by electron doping are represented as $\Delta n_\text{d}^\text{e} = \frac{1}{2}\frac{1}{L} \left[ \langle I \rangle_\text{doped} - \langle I \rangle_\text{half} \right]$, where $\langle \mathcal{O}\rangle_\text{doped}$ and $\langle \mathcal{O}\rangle_\text{half}$ are expectation values of $\mathcal{O}$ for electron-doped and half-filled 1DEHM, respectively.
The factor $1/2$ is introduced to compare the carrier density of electron-doped systems with that of pulse-excited systems where the same number of holons and doublons are excited.

\begin{figure}[t]
  \centering
    \includegraphics[clip, width=20pc]{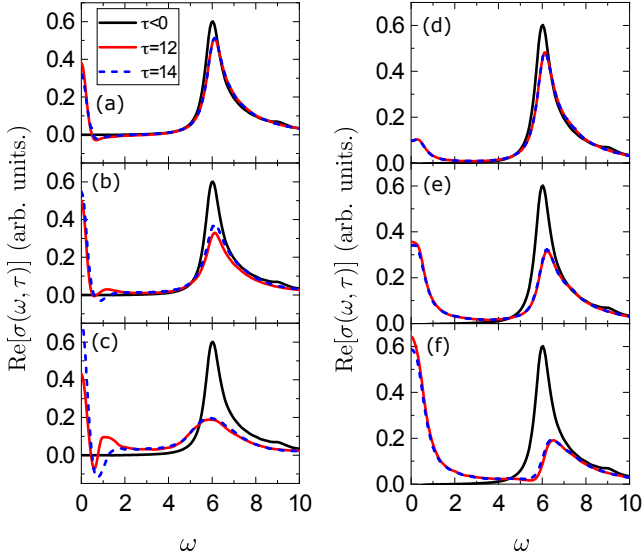}
    \caption{Re$[\sigma(\omega,\tau)]$ excited by $\Omega=3$ [$\Omega=6$] pulses with (a) $E_{0}=0.9$, (b) $E_{0}=1.5$, and (c) $E_{0}=1.8$ [(d) $E_{0}=0.12$, (e) $E_{0}=0.24$, and (f) $E_{0}=0.36$]. Black, red, and blue dashed lines are for $\tau<0$, $\tau=12$, and $14$, respectively. All plots are obtained by taking $\gamma=0.4$ for the half-filled $L=32$ 1DEHM with $(U,V)=(10,3)$.}
    \label{Fig4}
\end{figure}
We find that $\sigma_\text{D}$ of electron-doped 1DEHM (see Sec.~S2 in the Supplemental Material~\cite{Supplement}) has large values leading to $\sigma_\text{D}\propto \Delta n_\text{d}$.
Upon electron doping, electrons are free to move and their kinetic energy decreases as indicated by black points in Fig.~\ref{Fig3}(b).
The change of the kinetic energy for electron doped 1DEHM is defined as $\Delta T^\text{e}=-t_\text{h}\sum_{j,\sigma}\left[ \langle B_{j,\sigma}\rangle_\text{doped} - \langle B_{j,\sigma} \rangle_\text{half} \right]$.
Upon electron doping, spectral weights above the Mott gap transfer to those at $\omega=0$ due to spin-charge separation~\cite{Ogata1990}.
Since the change of total spectral weights is determined by $-\frac{1}{2L}\Delta T^\text{e}$ according to the optical sum rule~\cite{Maldague1977}, the decrease of kinetic energy contributes to the enhancement of $\sigma_\text{D}$.
Photon absorptions also lead to metallic states following $\overline{\sigma}_\text{D}\propto \Delta n_\text{d}$.
$\overline{\sigma}_\text{D}$ of 1DEHM excited by $\Omega=3$ and 6 pulses are obtained from Re$[\sigma(\omega,\tau)]$, which exhibits large spectral weights at $\omega = 0$ as shown in Fig.~\ref{Fig4}.
$\Omega=3$ pulses with $E_{0}=0.9$ [Fig.~\ref{Fig4}(a)], $E_{0}=1.5$ [Fig.~\ref{Fig4}(b)], and $E_{0}=1.8$ [Fig.~\ref{Fig4}(c)] lead to $\Delta n_{d}=0.007$, 0.03, and 0.07, respectively.
$\Omega=6$ pulses with $E_{0}=0.12$ [Fig.~\ref{Fig4}(d)], $E_{0}=0.24$ [Fig.~\ref{Fig4}(e)], and $E_{0}=0.36$ [Fig.~\ref{Fig4}(f)] lead to $\Delta n_{d}=0.009$, 0.03, and 0.05, respectively.
Note that $\overline{\sigma}_\text{D}$ is affected by the emergence of spectral weights at $\omega \sim 0.5$ (see Sec.~S3 in the Supplement Material~\cite{Supplement}).

In contrast to the electron-doped and photon-absorbed systems, there is no metallization when excitations are induced by a photon nonabsorbable $\Omega=0$ pulse causing quantum tunneling.
The change of kinetic energy $\Delta T=-t_\text{h}\sum_{j,\sigma}\left[ \overline{\langle B_{j,\sigma}\rangle_{t}} - \langle B_{j,\sigma} \rangle_{0} \right]$ induced by $\Omega=0$ pulses exhibits a significant difference from other cases: $-\Delta T<0$ monotonically decreases with increasing $\Delta n_\text{d}$ as shown by magenta and light blue points in Fig.~\ref{Fig3}(b).
We consider that a large increase in $\Delta T$ is associated with a restricted mobility due to the presence of strong fields, which leads to the strong suppression of $\overline{\sigma}_\text{D}$.

The time evolution of an entanglement entropy $S_\text{E}=-\sum_{i}p_{i}\ln p_{i}$ with the eigenvalue $p_{i}$ of a reduced density matrix obtained by contracting half of the whole system shows different behavior when 1DEHM is excited by quantum tunneling and by photon absorption (see Sec.~S4 in the Supplemental Material~\cite{Supplement}). 
For photon absorption, $S_\text{E}$ shows rapid linear growth and saturates at the end of pulse irradiation.
On the other hand, for quantum tunneling, $S_\text{E}$ shows slow logarithmic growth and continues to grow slowly even after the end of pulse irradiation. 
The slow growth of $S_\text{E}$~\cite{Znidaric2008, Bardarson2012, Serbyn2013, Vosk2013, Nanduri2014, Singh2016} is considered to be one of the manifestations of the localized nature of excited states by a high-field terahertz pulse.

\textit{Floquet effective Hamiltonians.}~We see how $\Omega=0$ pulses localize nonequilibrium states in the 1D Mott insulator.
For simplicity, we consider the dc limit of the Hamiltonian~(\ref{Eq-Hamiltonian}) with $V=0$ taking $A(t)=\Delta t$.
Using the Schrieffer-Wolff transformation~\cite{Bukov2015, Bukov2016}, we obtain an effective model for resonant driving $U=p\Delta \gg t_\text{h}$, taking non-zero integers $p$.
Due to the collective nature of the Zener breakdown, tunneling occurs not only between nearest-neighbor sites but also across several sites associated with $\Delta \leq U$~~\cite{Oka2003, Oka2005, Oka2008, Oka2010, Eckstein2010, Oka2012}.
The $\xi \sim1.5$ indicates that the dominant contribution to the breakdown is quantum tunneling within a few sites, which can be described as the effect of resonant electric fields with $\Delta=U/p$ for $p\lesssim3$.
The leading-order effective Hamiltonians for $p=1$, 2, and 3 are 
\begin{align*}
\mathcal{H}_{p=1}^{(0)}=&-t_\text{h}\sum_{j,\sigma}\left( h_{j+1j,\sigma}^{\dag} + h_{j+1j,\sigma} \right),\\
\mathcal{H}_{p=2}^{(1)}=&\frac{t_\text{h}^{2}}{\Delta}\left[
(T_{1}+T_{1}^{\dag})-2(T_{2}+T_{2}^{\dag})+H_{D}^{a}-T_{XY}
\right] \\
&+\frac{t_\text{h}^{2}}{3\Delta}\left(
H_{D}^{b}-T_{3}^{b}-T_{XY}
\right),\\
\mathcal{H}_{p=3}^{(1)}=&\frac{t_\text{h}^{2}}{2\Delta}\left(
H_{D}^{a}-T_{XY}
\right)
+\frac{t_\text{h}^{2}}{4\Delta}\left(
H_{D}^{b}-T_{3}^{b}-T_{XY}
\right),
\end{align*}
respectively (see Sec.~S5 in the Supplemental Material~\cite{Supplement}), where
\begin{align*}
h_{ji,\sigma}^{\dag} =& n_{j,-\sigma} (1-n_{i,-\sigma}) c_{j,\sigma}^{\dag} c_{i,\sigma},\\
T_{1}=& \sum_{j,\sigma} 
n_{j+2,-\sigma}(1-n_{j,-\sigma})(1-2n_{j+1,-\sigma})  c_{j+2,\sigma}^{\dag}c_{j,\sigma}, \\
T_{2}=& \sum_{j,\sigma} n_{j+2,\sigma}(1-n_{j,-\sigma})c_{j+2,-\sigma}^{\dag}c_{j+1,-\sigma}c_{j+1,\sigma}^{\dag}c_{j,\sigma},\\
H_{D}^{a}=&\sum_{j,\sigma} n_{j+1,-\sigma} \bigl[ - n_{j,\sigma} +2n_{j+1,\sigma} (1-n_{j,-\sigma}) \bigr],\\
H_{D}^{b}=&\sum_{j,\sigma} n_{j,\sigma}\bigl[ - n_{j+1,-\sigma} +2n_{j,-\sigma}(1-n_{j+1,-\sigma})  \bigr], \\
T_{3}^{b}=&\sum_{j,\sigma} n_{j,\sigma}\left(1 -n_{j+2,-\sigma} \right) \\
&\times \left( c_{j,-\sigma}c_{j+1,-\sigma}^{\dag} c_{j+1,\sigma}^{\dag} c_{j+2,\sigma} +\text{H.c.} \right),\\
T_{XY}=&\sum_{j,\sigma}  \left[(1-n_{j,-\sigma})(1-n_{j,\sigma}) + n_{j+1,-\sigma}n_{j+1,\sigma}\right] \\
&\times c_{j,-\sigma}^{\dag}c_{j+1,-\sigma}c_{j+1,\sigma}^{\dag}c_{j,\sigma}.
\end{align*}

The effective Hamiltonians suggest that the Floquet metastable states have conservations due to $\left[P+I,\mathcal{H}_{p=1}^{(0)} \right]=\left[P+2I,\mathcal{H}_{p=2}^{(1)} \right]=\left[P,\mathcal{H}_{p=3}^{(1)} \right]=\left[I,\mathcal{H}_{p=3}^{(1)} \right]=0$, where $P=\sum_{k}kn_{k}$ is the dipole moment.
Since the resonance condition induces real excitations, the effect of a strong electric field remains in excited states even after a pulse disappears.
Such conservation may break ergodicity and lead to exotic many-body dynamics.
Indeed, it has numerically demonstrated that $\mathcal{H}_{p=1}^{(0)}$ can induce ergodicity-breaking many-body eigenstates~\cite{Desaules2021} like quantum many-body scarring~\cite{Shiraishi2017, Moudgalya2018a, Moudgalya2018b, Iadecola2019a, Iadecola2019b, Ok2019, Turner2018a, Turner2018b, Choi2019, Lin2019, Ho2019, Feldmeier2019, Moudgalya2020}.
Also, dynamics governed by $\mathcal{H}_{p=2}^{(1)}$ is known to be non-ergodic~\cite{Scherg2021}.
Kinetic constraints imposed by such conservation lead to the emergent fragmentation of the Hilbert space, generating exponentially many disconnected subspaces~\cite{Moudgalya2019, Rakovszky2020, Khemani2020, Sala2020, Scherg2021, Desaules2021, Herviou2021, Kohlert2021, Papic2021, Moudgalya2021} even within a single symmetry sector.
Dipole-moment-conserved system is a representative system with such restriction as seen in fractons~\cite{Pretko2017a, Pretko2017b, Pretko2018, Williamson2019, Sous2020a, Sous2020b, Pretko2020, Nandkishore2019}, which localize charge excitations topologically.
$T_{3}^{b}$ included in $\mathcal{H}_{p=2}^{(1)}$ and $\mathcal{H}_{p=3}^{(1)}$ conserving both $P$ and $I$ is an example of showing doublon-assisted dipole-moment conserving processes, which does not produce Drude weight/superfluid density~\cite{Seidel2005}.

A strong $\Omega=0$ pulse produce two effects in excited states: one is the injection of carriers promoting itinerancy, and the other is the restriction of motion promoting localization.
As a result of their competing effects, the localization effect prevails in the $U\sim10$ strong coupling region, and the excited states follow glassy dynamics~\cite{Amir2011, Gopalakrishnan2014, vanHorssen2015, Prem2017, Pretko2017c, Lan2018} with weak-ergodicity breaking.
We see the strong suppression of $\sigma_\text{D}$ for $U=7$ and $13$ fixing $V/U=0.3$ (see Sec.~S6 in the Supplemental Material~\cite{Supplement}). 
However, the suppression of $\sigma_\text{D}$ for $U=7$ is weaker than that for $U=10$ and $13$.
This is because glassy states are unlikely to emerge in weak-coupling region, since the above discussion with the effective Hamiltonians is valid in strong-coupling regime.
We note that the glassy state proposed in this Letter has a different origin from that induced by randomness near the Mott transition~\cite{Dobrosavljevic2003, Dagotto2005, Miranda2005, Andrade2009, Itou2017, Yamamoto2020}.
We expect that the glassy dynamics may be detected in ET-F$_{2}$TCNQ excited by a terahertz pulse with amplitude about 3.5~MV/cm.

\textit{Summary.}~We have investigated Re$[\sigma(\omega,\tau)]$ of pulse-excited states of the half-filled 1DEHM using tDMRG.
We have proposed that an insulator-to-glass transition is induced by strong mono- and half-cycle pulses, which leads to the suppression of $\sigma_\text{D}$.
This is in contrast to the insulator-to-metal transition that occurs upon excitation by photon absorption accompanying $\sigma_\text{D} \propto \Delta n_\text{d}$.
Restricted mobility due to strong fields induces glassy dynamics as seen in fracton systems.
Not glassy but metallic states have been observed in the Mott insulator $\kappa$-(ET)$_{2}$Cu[N(CN)$_{2}$]Br excited by terahertz pulses in the experiment~\cite{Yamakawa2017}.
One possibility is that the enhancement of $\sigma_\text{D}$ has been observed during electric field irradiation when non-equilibrium metastable states have not yet been reached (see Sec.~S7 in the Supplemental Material~\cite{Supplement}).
Another possibility is that electron correlation is not so large that the subspaces in the fragmented Hilbert space are connected. 
Lastly, we note that qualitative differences in Re$[\sigma(\omega,\tau)]$ between photon absorptions and quantum tunnelings have recently been observed in a Mott insulator Ca$_2$RuO$_4$~\cite{Li2022}.

%%%%%%%%%%%%%%%%%%%%
% Acknowledgements
%%%%%%%%%%%%%%%%%%%%
\begin{acknowledgments}
We acknowledge discussions with H. Okamoto, K. Iwano, T. Yamaguchi, A. Takahashi, and Y. Murakami. 
This work was supported by CREST (Grant No. JPMJCR1661), the Japan Science and Technology Agency, by the Japan Society for the Promotion of Science, KAKENHI (Grants No. 17K14148, No. 19H01829, No. 19H05825, No. 21H03455) from Ministry of Education, Culture, Sports, Science, and Technology (MEXT), Japan, and by JST PRESTO (Grant No. JPMJPR2013). 
Numerical calculation was carried out using computational resources of HOKUSAI at RIKEN Advanced Institute for Computational Science, the supercomputer system at the information initiative center, Hokkaido University, the facilities of the Supercomputer Center at Institute for Solid State Physics, the University of Tokyo, and supercomputer Fugaku provided by the RIKEN Center for Computational Science through the HPCI System Research Project (Project ID: hp170325, hp220048).
\end{acknowledgments}

%
% Supplemental
%
\clearpage
\onecolumngrid

\begin{center}
{\bf \large Supplemental Material for ``Glassy dynamics of the one-dimensional Mott insulator excited by a strong terahertz pulse''}
\end{center}
\begin{center}
Kazuya Shinjo$^{1,2}$, Shigetoshi Sota$^{3}$, and Takami Tohyama$^1$\\
\vspace*{0.1cm}
{\footnotesize
$^1$Department of Applied Physics, Tokyo University of Science, Tokyo 125-8585, Japan\\
$^{2}$Computational Quantum Matter Research Team, RIKEN Center for Emergent Matter Science (CEMS), Wako, Saitama 351-0198, Japan\\
$^3$Computational Materials Science Research Team,
RIKEN Center for Computational Science (R-CCS), Kobe, Hyogo 650-0047, Japan
}
\end{center}

\renewcommand{\thesection}{S\arabic{section}}
\renewcommand{\theequation}{S\arabic{equation}}
\setcounter{equation}{0}
\renewcommand\thefigure{S\arabic{figure}}
\setcounter{figure}{0}
\renewcommand{\bibnumfmt}[1]{[S#1]}
\renewcommand{\citenumfont}[1]{S#1}

\section{Time-dependent optical conductivity in the nonequilibrium state}

Using a method discussed in Refs.~\cite{Shao2016, Shinjo2018}, we obtain optical conductivities in the nonequilibrium system. 
In order to identify the response of a system with respect to later probe pulses, subtraction is necessary; that is, two successive steps are involved in order to calculate the optical conductivity in nonequilibrium. 
First, a time-evolution process that describes the nonequilibrium development of a system in the absence of a probe pulse is evaluated. 
This leads to $ j_\text{pump}(t)$.
Second, in the presence of a probe pulse, we obtain $j_\text{total}(t,\tau)$. 
The subtraction of $j_\text{pump}(t)$ from $j_\text{total}(t,\tau)$ produces the required $j_\text{probe}(t,\tau)$, i.e., the variation of the current expectations due to the presence of a probe pulse.
Then, the optical conductivity in nonequilibrium is given by
\begin{align}
\sigma(\omega,\tau) = \frac{j_\text{probe}(\omega,\tau) }{i(\omega +i\eta)LA_\text{probe}(\omega)},
\end{align}
where $A_{\text{probe}}(\omega)$ and $j_\text{probe}(\omega,\tau)$ are the Fourier transform of the vector potential of a probe pulse $A_\text{probe}(t)$ and $j_\text{probe}(t,\tau)$, respectively.
$L$ is the number of sites.
This approach has been combined with DMRG to calculate optical conductivities~\cite{Ohmura2019, Shinjo2021, Shinjo2021b, Rincon2021}.

\begin{figure}[t]
  \centering
    \includegraphics[clip, width=20pc]{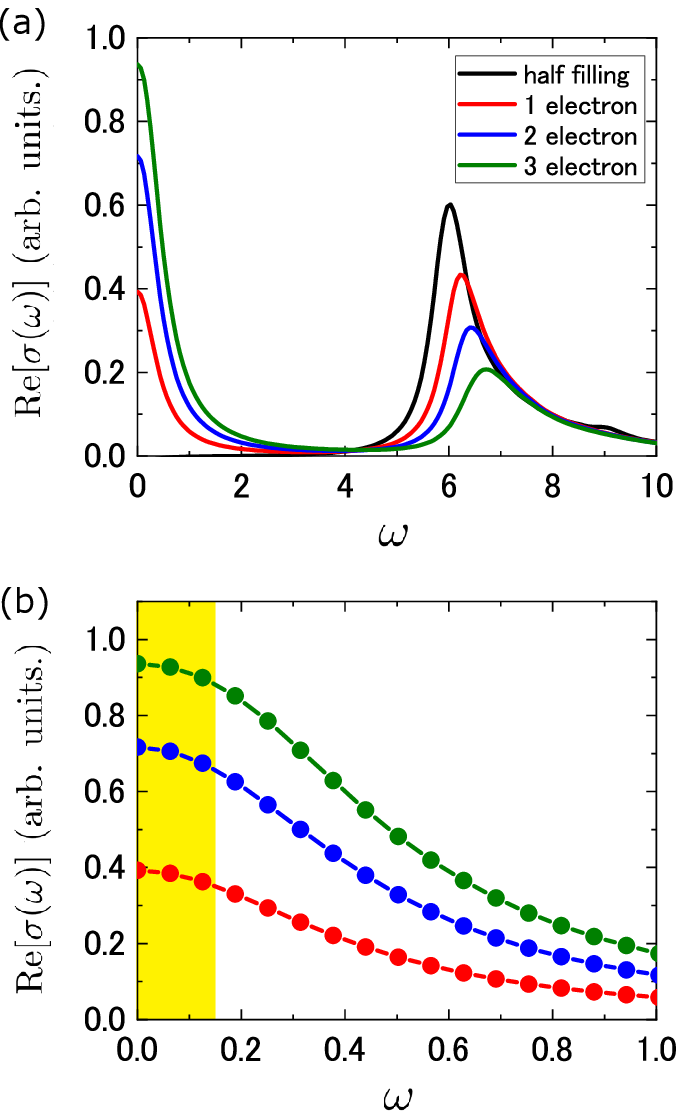}
    \caption{(a) Re$[\sigma(\omega)]$ of the $L=32$ 1DEHM with $(U,V)=(10,3)$, where we take $\gamma=0.4$. (b) The magnified view of the range $0\le \omega \le 1$.}
    \label{FigS1}
\end{figure}
\section{Optical conductivities in electron-doped systems}
We show in Fig.~\ref{FigS1}(a) the optical conductivities Re$[\sigma(\omega)]$ of the one-dimensional extended Hubbard model (1DEHM)
\begin{align}\label{Eq-Hamiltonian}
\mathcal{H}=&-t_\mathrm{h}\sum_{i,\sigma} \left(c_{i,\sigma}^\dag c_{i+1,\sigma}+\text{H.c.}\right)
+ U\sum_i n_{i,\uparrow}n_{i,\downarrow} + V\sum_i n_{i}n_{i+1}
\end{align}
with $(U,V)=(10,3)$ upon electron doping, where $c_{i,\sigma}^{\dag}$ is the creation operator of an electron with spin $\sigma (= \uparrow, \downarrow)$ at site $i$ and $n_i=\sum_\sigma n_{i,\sigma}$ with $n_{i,\sigma}=c^\dagger_{i,\sigma}c_{i,\sigma}$.
The black, red, blue, and green lines are for half-filling, one-, two-, and three-electron doping, respectively.
We use $L=32$ sites under open boundary conditions and a broadening factor $\gamma=0.4$.
Figure~\ref{FigS1}(b) shows the magnified view of the range $0\le \omega \le 1$.
We assume that spectral weights integrated in the yellow background $\sigma_\text{D}=\int_{\omega=0}^{2\eta}d\omega \text{Re}\sigma(\omega)$ with $2\eta=0.15$ contributes to the Drude weight.

\section{Metallic states generated by photon absorption}
We demonstrate that a metallic state with large $\sigma_\text{D}$ is induced in 1DEHM via photon absorption. In addition, the contribution of stimulated emission (SE) and absorption (SA) emerges since the transition dipole moment $\langle 1|x|2\rangle$ and a third-order nonlinear optical susceptibility are anomalously large in the 1D Mott insulators~\cite{Mizuno2000, Kishida2000, Tohyama2001, Kishida2001,Ono2004}, where $|1\rangle$ ($|2\rangle$) is one-photon-allowed (-forbidden) excitonic state with odd- (even-)parity symmetry. For $(U,V)=(10,3)$, the energy gap between $|0\rangle$ and $|1\rangle$ is $\omega_{1}=6$ and that between $|0\rangle$ and $|2\rangle$ is $\omega_{2}=6.3$, where $|0\rangle$ is the ground state.

\begin{figure}[t]
  \centering
    \includegraphics[clip, width=30pc]{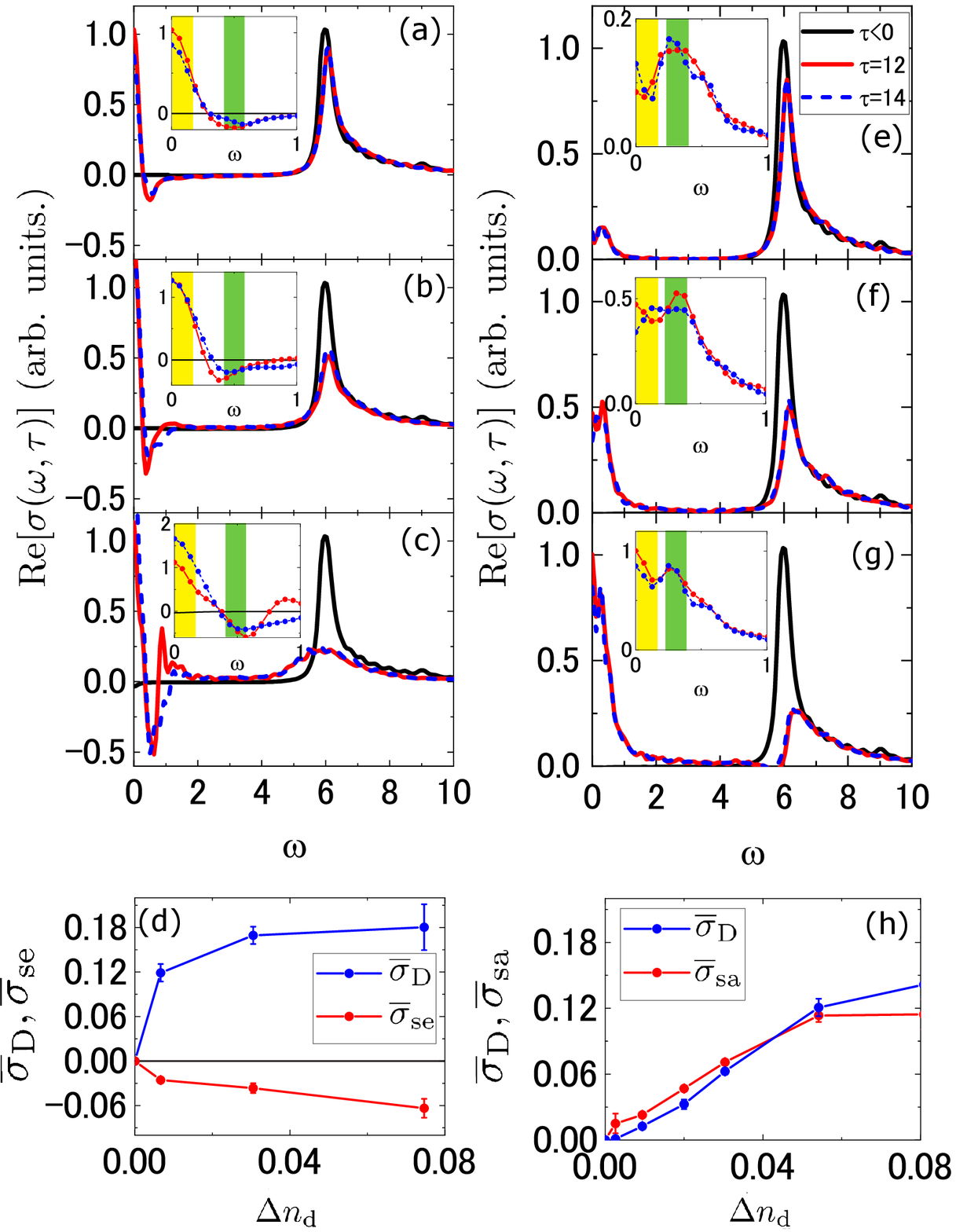}
    \caption{Re$[\sigma(\omega,\tau)]$ excited by $\Omega=3$ ($\Omega=6$) pulses with (a) $E_{0}=0.9$, (b) $E_{0}=1.5$, and (c) $E_{0}=1.8$ [(e) $E_{0}=0.12$, (f) $E_{0}=0.24$, and (g) $E_{0}=0.36$]. Black, red, and blue dashed lines are for $\tau<0$, $\tau=12$, and $14$, respectively. The insets show magnified views of the range $0\leq \omega \leq1$. (d) [(h)] $\overline{\sigma}_\text{D}$ and $\overline{\sigma}_\text{se}$ [$\overline{\sigma}_\text{sa}$] obtained from spectral weights in the yellow and green background in (a)-(c) [(e)-(g)]. Two endpoints of the error bars in (d) and (h) represent the value of $\sigma_\text{D}$ at $\tau=12$ and 14. All plots are obtained by taking $\gamma=0.2$ for the half-filled $L=32$ 1DEHM with $(U,V)=(10,3)$.}
    \label{FigS2}
\end{figure}
We show Re$[\sigma(\omega,\tau)]$ excited by $\Omega=3$ pulses with $E_{0}=0.9$ [Fig.~\ref{FigS2}(a)], 1.5 [Fig.~\ref{FigS2}(b)], and 1.8 [Fig.~\ref{FigS2}(c)], which leads to $\Delta n_{d}=0.007$, 0.03, and 0.07, respectively.
Here, $E_{0}$ is the amplitude of electric field and $\Delta n_\text{d}=\frac{1}{L}\bigl[ \overline{\langle I \rangle_{t}} - \langle I\rangle_0 \bigr]$ is the change in doublon density of 1DEHM before and after a pulse is applied, where $I=\sum_{j}n_{j,\uparrow}n_{j,\downarrow}$, $ \overline{\langle \mathcal{O} \rangle_{t}}$ is the average of an expectation value of an operator $\mathcal{O}$ from $t=21$ to $22$ just before a probe pulse is applied, and $\langle \mathcal{O} \rangle_0$ is an expectation value of $\mathcal{O}$ for a ground state.
$\Omega=3$ pulses excite the ground state to $|2\rangle$ by a two-photon process.
$\sigma_\text{D}$ has finite values even for open boundary conditions, since we introduce $\gamma=0.2$ in Fig.~\ref{FigS2}.
Figures~\ref{FigS2}(a)-(c) and \ref{FigS2}(e)-(g) are the same as Figs.~4(a)-(c) and 4(d)-(f) in the main text, but $\gamma$ is changed.
To distinguish the fine structure of spectra, $\gamma$ used in this section is smaller than that in other sections and the main text.
In addition to metallic properties following $\sigma_\text{D}\propto \Delta n_\text{d}$, we find negative spectral weights at $\omega \simeq 0.5$ due to SE from $|2\rangle$ to $|1\rangle$.
Since $\sigma_\text{D}$ has large values, the energy $\omega \simeq 0.5$ that takes a minimum of Re$[\sigma(\omega,\tau)]$ is different from $\delta \omega =\omega_{2}-\omega_{1}= 0.3$.
We show in Fig.~\ref{FigS2}(d) spectral weights contributing to $\sigma_\text{D}$ and SE estimated as $\overline{\sigma}_\text{D}=\frac{1}{2}\sum_{\tau=12,14}\int_{\omega=0}^{2\eta}d\omega \text{Re}\sigma(\omega,\tau)$ with blue points and $\overline{\sigma}_\text{se}=\frac{1}{2}\sum_{\tau=12,14}\int_{\omega=0.5-\eta}^{0.5+\eta} d\omega \text{Re}\sigma(\omega,\tau)$ with red points, respectively, where the energy width is defined as $2\eta=0.15$.
We find that $\overline{\sigma}_\text{D}>0$ ($\overline{\sigma}_\text{se}<0$) increases (decreases) with increasing $\Delta n_\text{d}$.
We note here that the redshift of the Mott gap for $\Delta n_\text{d}=0.07$ as shown in Fig.~\ref{FigS2}(c) is due to the dynamical Coulomb screening~\cite{Golez2015, Baykusheva2022}.

We show Re$[\sigma(\omega,\tau)]$ excited by $\Omega=6$ pulses with $E_{0}=0.12$ [Fig.~\ref{FigS2}(e)], $0.24$ [Fig.~\ref{FigS2}(f)], and $0.36$ [Fig.~\ref{FigS2}(g)], which leads to $\Delta n_{d}=0.009$, 0.03, and 0.05, respectively.
These pulses excite the ground state to $|1\rangle$ by a one-photon process.
As with the case of $\Omega=3$ pulses, metallic states are induced accompanying $\sigma_\text{D}\propto \Delta n_\text{d}$.
However, we do not find negative spectral weights but positive ones at $\omega \simeq \delta \omega$ due to SA from $|1\rangle$ to $|2\rangle$~\cite{Shao2016, Shinjo2018, Rincon2021}.
We show in Fig.~\ref{FigS2}(h) spectral weights contributing to $\sigma_\text{D}$ and SA estimated as $\overline{\sigma}_\text{D}$ with blue points and $\overline{\sigma}_\text{sa}=\frac{1}{2}\sum_{\tau=12,14}\int_{\omega=0.3-\eta}^{0.3+\eta}d\omega \text{Re}\sigma(\omega,\tau)$ with red points, respectively.
We find that both $\overline{\sigma}_\text{D}$ and $\overline{\sigma}_\text{sa}$ increase with increasing $\Delta n_\text{d}$.

\section{The time evolution of Entanglement entropy}
\begin{figure}[t]
  \centering
    \includegraphics[clip, width=30pc]{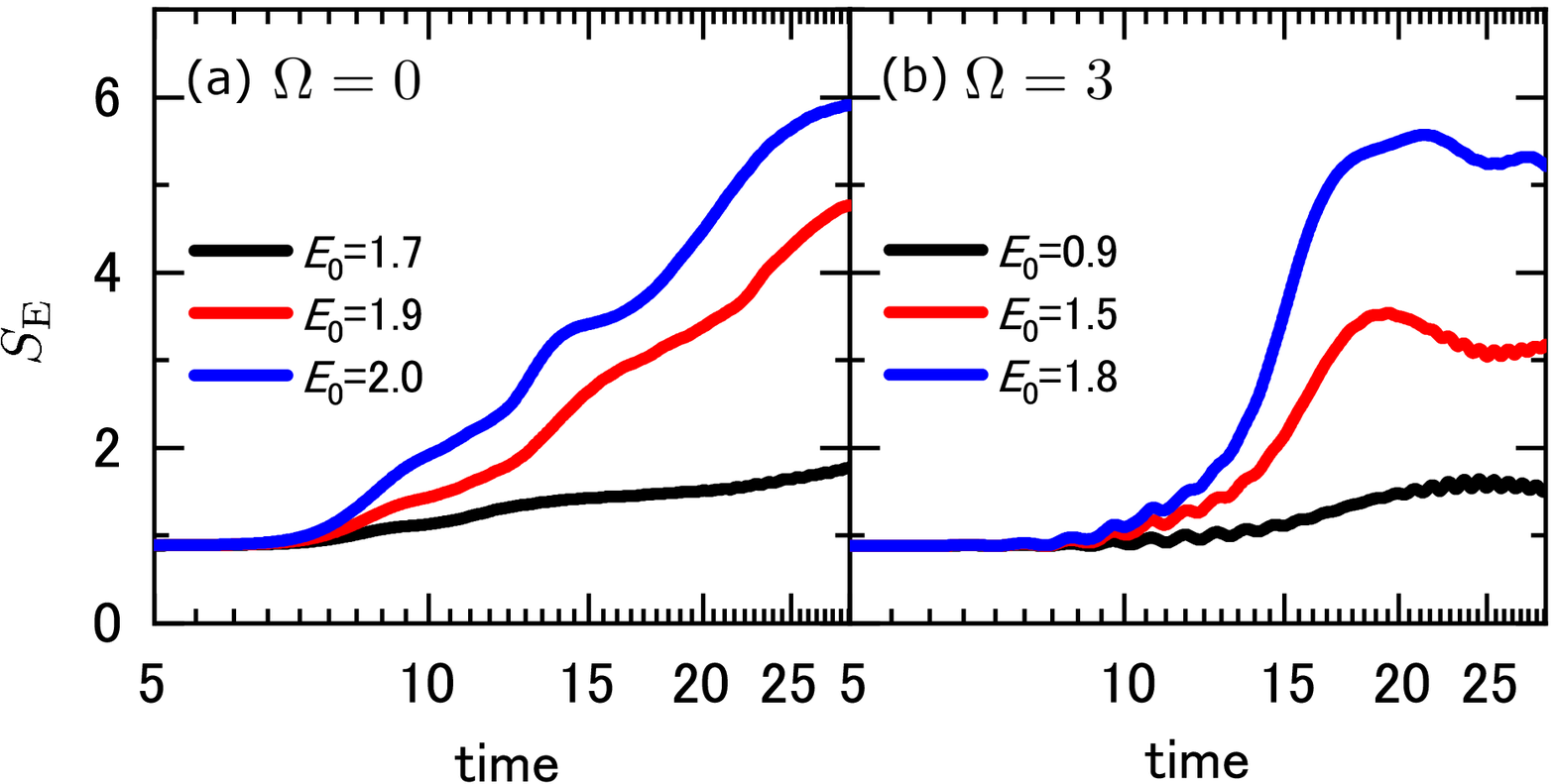}
    \caption{The time evolution of $S_\text{E}$ in the half-filled $L=32$ 1DEHM excited by (a) a $\Omega=0$ mono-cycle pulse and (b) a $\Omega=3$ pulse, which induces quantum tunneling and photon absorption, respectively.}
    \label{FigS3}
\end{figure}
The time evolution of an entanglement entropy $S_\text{E}=-\sum_{i}p_{i}\ln p_{i}$ shows different behavior when 1DEHM is excited by quantum tunneling and by photon absorption.
Here, $p_{i}$ is obtained by the Schmidt decomposition of a wavefunction $|\psi \rangle$ as
\begin{align}
|\psi \rangle = \sum_{i} p_{i} |\psi_A^{i}\rangle |\psi_B^{i}\rangle,
\end{align}
where a system is composed of two subsystems $A$ and $B$.
We obtain $S_\text{E}$ by making $A$ half of the whole system.
In Fig.~\ref{FigS3}(a), we show the time evolution of $S_\text{E}$ in $L=32$ half-filled 1DEHM excited by a $\Omega=0$ mono-cycle pulse, which induces quantum tunneling.
This pulse is the same as the one used in Figs.~1(d), 1(e), and 2(a)-2(c) in the main text.
We find that $S_\text{E}$ shows slow logarithmic growth and continues to grow slowly even after the end of pulse irradiation, i.e., $t>20$.
For comparison, we show in Fig.~\ref{FigS3}(b) the time evolution of $S_\text{E}$ when 1DEHM is excited with a $\Omega=3$ pulse, whose photons are absorbable.
This pulse is the same as the one used in Figs.~1(c) and 4(a)-4(c) in the main text.
We see that $S_\text{E}$ shows rapid linear growth and then saturates at the end of pulse irradiation, i.e., $t=20$.
Since the initial state is not a product state with $S_\text{E}= 0$, and the increase in $S_\text{E}$ depends on the amplitudes of electric fields $E_{0}$, the situation is complicated, and underlying physics is nontrivial.
However, we can see a tendency for $S_\text{E}$ to grow more slowly for quantum tunneling than for photon absorption.
The slow growth of $S_\text{E}$~\cite{Znidaric2008, Bardarson2012, Serbyn2013, Vosk2013, Nanduri2014, Singh2016} is considered to be one of the manifestations of the localized nature of excited states by a high-field terahertz pulse.
Since entanglement spectrum contains more information than $S_\text{E}$, its analysis is interesting and remains as a future work.

\section{Effective Hamiltonians with strong couplings and fields}
For a time-periodic Hamiltonian $\mathcal{H}(t+T_{0})=\mathcal{H}(t)$, we obtain the effective Hamiltonian in the high-frequency limit $\Omega_{0}=2\pi/T_{0} \gg 1$ as $\mathcal{H}_\text{eff}=\mathcal{H}_\text{eff}^{(0)} + \mathcal{H}_\text{eff}^{(1)} + O(\Omega_{0}^{-2})$~\cite{Eckardt2015, Bukov2015, Bukov2016}, where
\begin{align}
\mathcal{H}_\text{eff}^{(0)}=& \frac{1}{T_{0}}\int_{0}^{T_{0}} dt \mathcal{H}(t)=\mathcal{H}_{0}, \\
\mathcal{H}_\text{eff}^{(1)}=& \sum_{m=1}^{\infty} \frac{\left[ \mathcal{H}_{m},\mathcal{H}_{-m} \right]}{m\Omega_{0}}.
\end{align}
Here, we Fourier-decompose the Hamiltonian as $\mathcal{H}(t)=\sum_{l=-\infty}^{\infty}\mathcal{H}_{l}e^{il\Omega_{0}t}$.

We consider the one-dimensional Hubbard model with dc electric field $\Delta$
\begin{align}
\mathcal{H}_{\Delta} = \sum_{j,\sigma}\left[-t_\text{h} \left(c_{j,\sigma}^\dag c_{j+1,\sigma}+\text{H.c.}\right) +\Delta j n_{j,\sigma} \right] +U\sum_{j} n_{j,\uparrow}n_{j,\downarrow}.
\end{align}
$\mathcal{H}_{\Delta}$ is represented in the rotating frame as
\begin{align}
H_{\text{rot}}(t)
=&-t_\text{h}\sum_{j,\sigma} \Bigl[e^{i\Delta t}g_{j j+1,\sigma} + e^{-i\Delta t}g_{jj+1,\sigma}^{\dag}\nonumber \\
&+ e^{i(\Delta+U)t} h_{jj+1,\sigma}^{\dag} + e^{-i(\Delta+U)t} h_{jj+1,\sigma} + e^{-i(\Delta-U)t} h_{j+1j,\sigma}^{\dag} + e^{i(\Delta-U)t} h_{j+1j,\sigma} \Bigr] \end{align}
with respect to the rotating operator 
\begin{align}
V(t)= e^{-it\left[\Delta \sum_{j,\sigma} jn_{j,\sigma} + U\sum_j n_{j,\uparrow}n_{j,\downarrow}\right]}.
\end{align}
Here, we define
\begin{align}
g_{ij,\sigma} =& g_{ij,\sigma}^{\dag} = (1-n_{i,-\sigma})(1-n_{j,-\sigma})c_{i,\sigma}^{\dag}c_{j,\sigma} +  n_{i,-\sigma} n_{j,-\sigma}c_{i,\sigma}^{\dag}c_{j,\sigma}, \\
h_{ij,\sigma}^{\dag} =& n_{i,-\sigma}(1-n_{j,-\sigma})c_{i,\sigma}^{\dag}c_{j,\sigma}.
\end{align}
For $U=p\Delta$ with non-zero intergers $p$, we obtain
\begin{align}
H_{\text{rot}}(t)
=&-t_\text{h}\sum_{j,\sigma} \Bigl[e^{i\Delta t}g_{jj+1,\sigma} +e^{-i\Delta t}g_{jj+1,\sigma}^{\dag}\nonumber \\
&+ e^{i(p+1)\Delta t} h_{jj+1,\sigma}^{\dag}  + e^{-i(p+1)\Delta t} h_{jj+1,\sigma} + e^{i(p-1)\Delta t} h_{j+1j,\sigma}^{\dag} + e^{-i(p-1)\Delta t} h_{j+1j,\sigma} \Bigr].
\end{align}

\subsection{The case where $p=1$}
For $p=1$, the Fourier-decomposed Hamiltonian reads
\begin{align}
H_{\text{rot}}(t)
= \sum_{m=-\infty}^{\infty} e^{im\Delta t} H_{t,m}^{p=1},
\end{align}
with
\begin{align}
H_{t,0}^{p=1} =& -t_\text{h}\sum_{j ,\sigma} \left[ h_{j+1j,\sigma}^{\dag} + h_{j+1j,\sigma}\right], \\
H_{t,1}^{p=1} =&-t_\text{h}\sum_{j,\sigma} g_{jj+1,\sigma},\\
H_{t,-1}^{p=1} =&-t_\text{h}\sum_{j,\sigma} g_{jj+1,\sigma}^{\dag}, \\
H_{t,2}^{p=1} =&-t_\text{h}\sum_{j,\sigma} h_{jj+1,\sigma}^{\dag}, \\
H_{t,-2}^{p=1} =&-t_\text{h}\sum_{j,\sigma} h_{jj+1,\sigma},
\end{align}
and
\begin{align}
H_{t,m}^{p=1} = 0 \text{    (} |m|>2).
\end{align}
With high-frequency expansion, we obtain the leading-order effective Hamiltonian as
\begin{align}
\mathcal{H}_{p=1}^{(0)}=\mathcal{H}_\text{eff}^{(0)}=H_{t,0}^{p=1}.
\end{align}
$\mathcal{H}_{p=1}^{(0)}$ has a conservation due to
\begin{align}
\left[ \Delta \sum_{k}kn_{k} +U \sum_{k}n_{k,\uparrow}n_{k,\downarrow} ,\mathcal{H}_{p=1}^{(0)} \right]=&
0
\end{align}
and has been studied in Ref.~\cite{Desaules2021}, which proposes the realization of quantum many-body scars.

\subsection{The case where $p=2$}
For $p=2$, the Fourier-decomposed Hamiltonian reads
\begin{align}
H_{\text{rot}}(t)
= \sum_{m=-\infty}^{\infty} e^{im\Delta t} H_{t,m}^{p=2},
\end{align}
where 
\begin{align}
H_{t,0}^{p=2} =& 0, \\
H_{t,1}^{p=2} =&-t_\text{h}\sum_{j,\sigma} \left(g_{jj+1,\sigma} + h_{j+1j,\sigma}^{\dag} \right), \\
H_{t,-1}^{p=2} =&-t_\text{h}\sum_{j,\sigma} \left( g_{jj+1,\sigma}^{\dag} + h_{j+1j,\sigma} \right), \\
H_{t,2}^{p=2} =&0, \\
H_{t,-2}^{p=2} =&0, \\
H_{t,3}^{p=2} =&-t_\text{h}\sum_{j,\sigma} h_{jj+1,\sigma}^{\dag}, \\
H_{t,-3}^{p=2} =&-t_\text{h}\sum_{j,\sigma} h_{jj+1,\sigma}, 
\end{align}
and
\begin{align}
H_{t,m}^{p=2} = 0 \text{    (} |m|>4).
\end{align}
With high-frequency expansion, we obtain the leading-order effective Hamiltonian as
\begin{align}
\mathcal{H}_{p=2}^{(1)}=&\mathcal{H}_\text{eff}^{(1)}=\frac{[H_{t,1}^{p=2},H_{t,-1}^{p=2}]}{\Delta} + \frac{[H_{t,3}^{p=2},H_{t,-3}^{p=2}]}{3\Delta}\nonumber \\
=&\frac{t_\text{h}^{2}}{\Delta}\left[
(T_{1}+T_{1}^{\dag})-2(T_{2}+T_{2}^{\dag})+H_{D}^{a}-T_{3}^{a}-T_{XY}
\right]
+\frac{t_\text{h}^{2}}{3\Delta}\left(
H_{D}^{b}-T_{3}^{b}-T_{XY}
\right),
\end{align}
since
\begin{align}
\mathcal{H}_{p=2}^{(0)}=\mathcal{H}_\text{eff}^{(0)}=0.
\end{align}
Here, we use the commutation relations as follows.
\begin{align}
\left[ -t_\text{h}\sum_{j,\sigma} g_{jj+1,\sigma}, -t_\text{h}\sum_{j,\sigma} g_{jj+1,\sigma}^{\dag} \right]=0,
\end{align}

\begin{align}
\left[ -t_\text{h}\sum_{j,\sigma} h_{j+1j,\sigma}^{\dag}, -t_\text{h}\sum_{j,\sigma} g_{jj+1,\sigma}^{\dag} \right]
=t_\text{h}^{2}\left( T_{1} -2 T_{2} \right),
\end{align}
\begin{align}
\left[ -t_\text{h}\sum_{j,\sigma} g_{jj+1,\sigma}, -t_\text{h}\sum_{j,\sigma} h_{j+1j,\sigma} \right]
=t_\text{h}^{2}\left( T_{1}^{\dag} -2 T_{2}^{\dag} \right),
\end{align}
\begin{align}
\left[ -t_\text{h}\sum_{j,\sigma} h_{j+1j,\sigma}^{\dag}, -t_\text{h}\sum_{j,\sigma} h_{j+1j,\sigma} \right]
=t_\text{h}^{2}(H_{D}^{a}-T_{3}^{a}-T_{XY}),
\end{align}
\begin{align}
\left[ -t_\text{h}\sum_{j,\sigma} h_{jj+1,\sigma}^{\dag}, -t_\text{h}\sum_{j,\sigma} h_{jj+1,\sigma} \right]
=t_\text{h}^{2} \left( H_{D}^{b}-T_{3}^{b}-T_{XY}\right),
\end{align}
where
\begin{align}
T_{1}=& \sum_{j,\sigma} 
n_{j+2,-\sigma}(1-n_{j,-\sigma})(1-2n_{j+1,-\sigma})  c_{j+2,\sigma}^{\dag}c_{j,\sigma}, \\
T_{2}=& \sum_{j,\sigma} n_{j+2,\sigma}(1-n_{j,-\sigma})c_{j+2,-\sigma}^{\dag}c_{j+1,-\sigma}c_{j+1,\sigma}^{\dag}c_{j,\sigma},\\
H_{D}^{a}=&\sum_{j,\sigma} \left( - n_{j,\sigma}n_{j+1,-\sigma} +2n_{j,-\sigma}n_{j,\sigma} -2 n_{j,-\sigma}n_{j+1,-\sigma}n_{j+1,\sigma} \right),\\
H_{D}^{b}=&\sum_{j,\sigma} \left( - n_{j,\sigma}n_{j+1,-\sigma} +2n_{j,-\sigma}n_{j,\sigma} -2 n_{j,-\sigma}n_{j,\sigma}n_{j+1,-\sigma} \right), \\
T_{3}^{a}=&\sum_{j,\sigma} (1-n_{j,-\sigma})n_{j+2,\sigma} \left(c_{j,-\sigma}c_{j+1,-\sigma}^{\dag} c_{j+1,\sigma}^{\dag}c_{j+2,\sigma} +\text{H.c.}\right)=0, \\
T_{3}^{b}=&\sum_{j,\sigma} n_{j,\sigma}\left(1 -n_{j+2,-\sigma} \right) \left( c_{j,-\sigma}c_{j+1,-\sigma}^{\dag} c_{j+1,\sigma}^{\dag} c_{j+2,\sigma} +\text{H.c.} \right),\\
T_{XY}=&\sum_{j,\sigma}  \left[(1-n_{j,-\sigma})(1-n_{j,\sigma}) + n_{j+1,-\sigma}n_{j+1,\sigma}\right] c_{j,-\sigma}^{\dag}c_{j+1,-\sigma}c_{j+1,\sigma}^{\dag}c_{j,\sigma}.
\end{align}
For $\mathcal{H}_{p=2}^{(1)}$, we find
\begin{align}
\left[ \sum_{k}kn_{k}+2\sum_{k}n_{k,\uparrow}n_{k,\downarrow}, \mathcal{H}_{p=2}^{(1)} \right]=0.
\end{align}
For initial states with charge-density waves, non-ergodic dynamics has been proposed in the case of $p=2$~\cite{Scherg2021}.

\subsection{The case where $p=3$}
For $p=3$, the Fourier-decomposed Hamiltonian reads
\begin{align}
H_{\text{rot}}(t)
= \sum_{m=-\infty}^{\infty} e^{im\Delta t} H_{t,m}^{p=3},
\end{align}
where 
\begin{align}
H_{t,0}^{p=3} =& 0, \\
H_{t,1}^{p=3} =&-t_\text{h}\sum_{j,\sigma} g_{jj+1,\sigma},\\
H_{t,-1}^{p=3} =&-t_\text{h}\sum_{j,\sigma} g_{jj+1,\sigma}^{\dag}, \\
H_{t,2}^{p=3} =&-t_\text{h}\sum_{j,\sigma} h_{j+1j,\sigma}^{\dag}, \\
H_{t,-2}^{p=3} =&-t_\text{h}\sum_{j,\sigma} h_{j+1j,\sigma}, \\
H_{t,3}^{p=3} =&0,\\
H_{t,-3}^{p=3} =&0,\\
H_{t,4}^{p=3} =&-t_\text{h}\sum_{j,\sigma} h_{jj+1,\sigma}^{\dag}, \\
H_{t,-4}^{p=3} =&-t_\text{h}\sum_{j,\sigma} h_{jj+1,\sigma},
\end{align}
and
\begin{align}
H_{t,m}^{p=3} = 0 \text{    (} |m|>5).
\end{align}
With high-frequency expansion, we obtain the leading-order effective Hamiltonian as
\begin{align}
\mathcal{H}_{p=3}^{(1)}=&\mathcal{H}_\text{eff}^{(1)}=\frac{[H_{t,1}^{p=3},H_{t,-1}^{p=3}]}{\Delta} + \frac{[H_{t,2}^{p=3},H_{t,-2}^{p=3}]}{2\Delta} + \frac{[H_{t,4}^{p=3},H_{t,-4}^{p=3}]}{4\Delta} \nonumber \\
=&\frac{t_\text{h}^{2}}{2\Delta}\left(
H_{D}^{a}-T_{3}^{a}-T_{XY}
\right)
+\frac{t_\text{h}^{2}}{4\Delta}\left(
H_{D}^{b}-T_{3}^{b}-T_{XY}
\right),
\end{align}
since
\begin{align}
\mathcal{H}_{p=3}^{(0)}=\mathcal{H}_\text{eff}^{(0)}=0.
\end{align}
For $\mathcal{H}_{p=3}^{(1)}$, we find
\begin{align}
\left[ \sum_{k}kn_{k}, \mathcal{H}_{p=3}^{(1)} \right]=
\left[ \sum_{k}n_{k,\uparrow}n_{k,\downarrow}, \mathcal{H}_{p=3}^{(1)} \right]=0.
\end{align}

\section{Interaction dependence of optical conductivities excited by a $\Omega=0$ pulse}
\begin{figure}[t]
  \centering
    \includegraphics[clip, width=30pc]{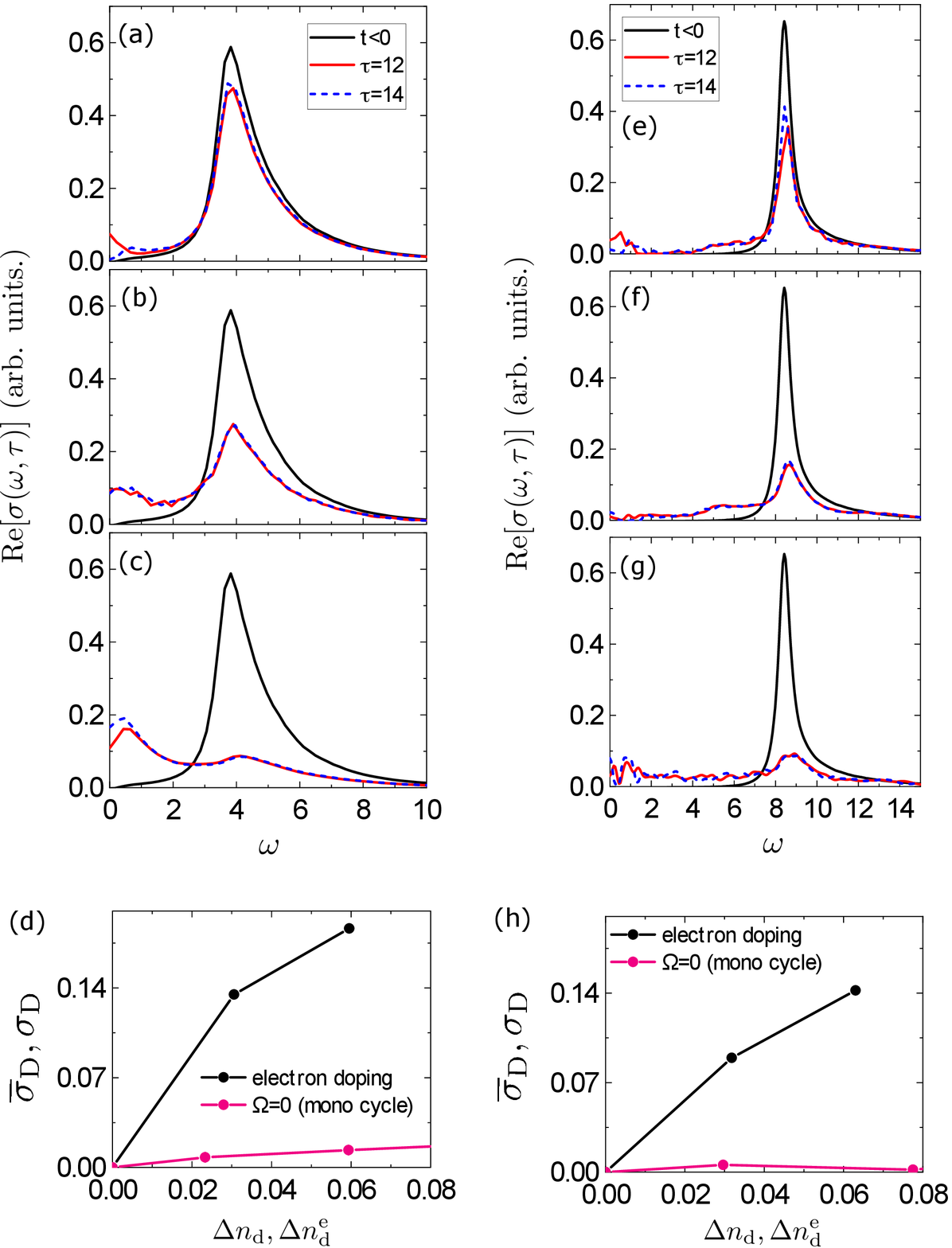}
    \caption{Re$[\sigma(\omega,\tau)]$ of the half-filled $L=32$ 1DEHM excited by $\Omega=0$ mono-cycle pulses with $t_\text{d}=2$ for (a)-(d) $(U,V)=(7,2.1)$ and (e)-(h) $(U,V)=(13,3.9)$. $\gamma=0.4$ is taken. The amplitudes of electric fields $E_{0}$ are (a) $E_{0}=0.9$, (b) $E_{0}=1.1$, (c) $E_{0}=1.4$, (e) $E_{0}=2.7$, (f) $E_{0}=2.9$, and (g) $E_{0}=3.0$. Black, red, and blue dashed lines are for $\tau<0$, $\tau=12$, and $14$, respectively. $\overline{\sigma}_\text{D}$ as a function of $\Delta n_\text{d}$ and $\Delta n_\text{d}^\text{e}$ for (d) $(U,V)=(7,2.1)$ and (h) $(U,V)=(13,3.9)$.}
    \label{FigS4}
\end{figure}
We show Re$[\sigma(\omega,\tau)]$ of 1DEHM excited by a $\Omega=0$ pulse for $U=7$ and $13$ in Figs.~\ref{FigS4}(a)-(c) and Figs.~\ref{FigS4}(e)-(g), respectively.
Here, we fix $V/U=0.3$.
The parameters for pump pulses $t_\text{d}$ and $t_{0}$ are the same as those used for Fig.~2 in the main text.
We obtain $\Delta n_\text{d}=0.02$, 0.06, and 0.1 for Figs.~\ref{FigS4}(a), \ref{FigS4}(b), and \ref{FigS4}(c), respectively.
Also, we obtain $\Delta n_\text{d}=0.03$, 0.08, and 0.1 for Figs.~\ref{FigS4}(e), \ref{FigS4}(f), and \ref{FigS4}(g), respectively.
From Figs.~\ref{FigS4}(a)-(c) and \ref{FigS4}(e)-(g), we can characterize the Drude weights $\overline{\sigma}_\text{D}$, which are shown as a function of $\Delta n_\text{d}$ and $\Delta n_\text{d}^\text{e}$ in Fig.~\ref{FigS4}(d) and \ref{FigS4}(h) for $(U,V)=(7,2.1)$ and $(U,V)=(13,3.9)$, respectively.
Here, carrier density by electron doping is $\Delta n_\text{d}^\text{e} = \frac{1}{2}\frac{1}{L} \left[ \langle I \rangle_\text{doped} - \langle I \rangle_\text{half} \right]$, where $\langle \mathcal{O}\rangle_\text{doped}$ and $\langle \mathcal{O}\rangle_\text{half}$ are expectation values of $\mathcal{O}$ for electron-doped and half-filled 1DEHM, respectively. 
Similar to the main text, we compare $\overline{\sigma}_\text{D}$ induced by chemical doping and a $\Omega=0$ pulse in Figs.~\ref{FigS4}(d) and \ref{FigS4}(h).
These figures indicate that $\sigma_\text{D}$ is strongly suppressed at $U=7$ and $13$ as well as for $U=10$ discussed in the main text.
We note that the suppression of $\sigma_\text{D}$ for $U=7$ is weaker than that for $U=10$ and 13.
The $U$ dependence of $\sigma_\text{D}$ indicates that glassy states are unlikely to emerge in the weak-coupling region.

\section{The examination of pump-pulse widths in optical spectra}
As in the main text, we consider Re$[\sigma(\omega,\tau)]$ of the half-filled 1DEHM excited by electric pulses whose vector potential is represented as $A_\text{pump}(t)=A_0 e^{-(t-t_0)^2/(2t_\mathrm{d}^2)} \cos\left[\Omega(t-t_0)\right]$.
We examine the case where 1DEHM is excited by $\Omega=0$ pulses with longer width than the ones used in the main text.
In this section, we use $(t_\text{d},t_{0})=(20,100)$, which leads to the central frequency 2.5~THz and time period 0.4~ps taking $t_\text{h}=112\text{meV}$ for ET-F$_{2}$TCNQ.
This condition corresponds to pulses used in the experiment~\cite{Yamakawa2017}, which are about 10 times wider than the one shown in the inset of Fig.~2(d) in the main text.
\begin{figure}[t]
  \centering
    \includegraphics[clip, width=16pc]{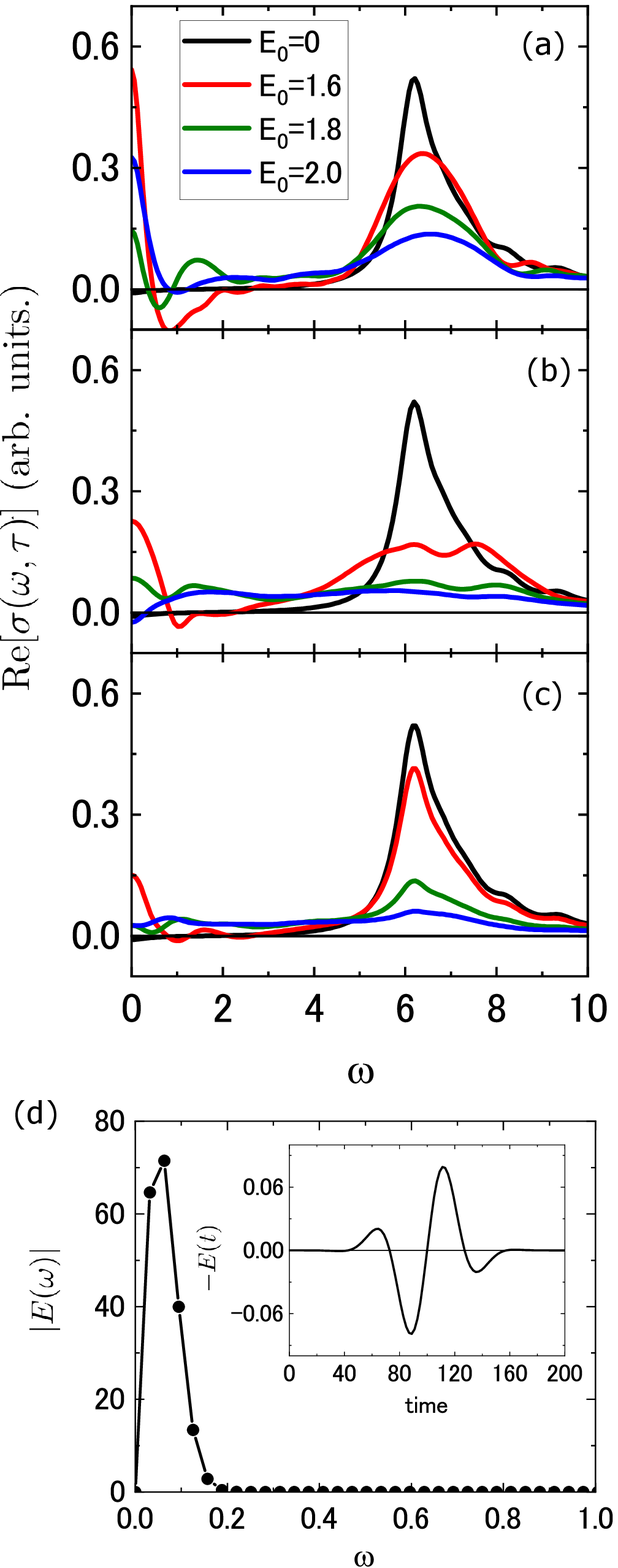}
    \caption{Re$[\sigma(\omega,\tau)]$ excited by $\Omega=0$ mono-cycle pulses for $t_\text{d}=20$ with (a) $\tau=0$, (b) $\tau=20$, and (c) $\tau=80$. (d) $|E(\omega)|$ with $E_{0}=0.08$. The inset indicates $-E(t)$.}
    \label{FigS5}
\end{figure}

We show Re$[\sigma(\omega,\tau)]$ for the three cases of (a) $\tau=0$, (b) $\tau=20$, and (c) $\tau=80$ in Figs.~\ref{FigS5}(a), \ref{FigS5}(b), and \ref{FigS5}(c), respectively.
As in the main text, probe pulses are applied at $t_{0}+\tau$.
We calculate Re$[\sigma(\omega,\tau)]$ of $L=12$ half-filled 1DEHM for $(U,V)=(10,3)$ with the time-dependent Lanczos method under open boundary conditions. 
As shown in Fig.~\ref{FigS5}(d), the spectral width of a pump pulse used in this section is narrower than that used in the main text as shown in Fig.~2(d) and 2(h).
Since $E(t)$ is given as shown in the inset of Fig.~\ref{FigS5}(d), Re$[\sigma(\omega,\tau)]$ with $\tau=0$ and 20 gives spectra near the center of applied pump pulses.
Whereas, Re$[\sigma(\omega,\tau=80)]$ gives spectra after turning off pump pulses, which is the same situation as the one considered in the main text.
The changes in doublon density between $t=0$ and 180 are 0.025, 0.093, and 0.12 for $E_{0}=1.6$, 1.8, and 2.0, respectively.
Since the virtual creation and annihilation of doublons and holons occur during the application of electric pulses, Re$[\sigma(\omega,\tau=0)]$ shown in Fig.~\ref{FigS5}(a) exhibits a complicated behavior and cannot be simply understood.
Nevertheless, we can see that $\sigma_\text{D}$ tends to have a large oscillating values.
$\sigma_\text{D}$ begins to be suppressed at $\tau=20$ and has small values at $\tau=80$ especially for $E_{0}\ge1.8$.
We find that the strong suppression of $\sigma_\text{D}$ captured in the main text appears again for $t_\text{d}=20$ if a probe pulse is applied after turning off pump pulses.

\end{document}